\documentclass[aip, pof, reprint]{revtex4-2}
\usepackage{amsfonts, amsmath, bm, amssymb, color}
\usepackage{graphicx}
\usepackage[colorlinks=true, citecolor=blue, linkcolor=blue]{hyperref}

\usepackage[T2A]{fontenc}
\usepackage[utf8]{inputenc}

\begin{document}

\title{Learning minimum-time navigation policies in two-dimensional flows with a genetic algorithm}

\author{Vladimir Parfenyev}\email{parfenius@gmail.com}
\affiliation{Landau Institute for Theoretical Physics of the Russian Academy of Sciences, 142432 Chernogolovka, Russia}
\affiliation{HSE University, 101000 Moscow, Russia}

\date{\today}

\begin{abstract}
We study minimum-time navigation between two points in a known two-dimensional fluid flow for a vessel with fixed-magnitude slip velocity and controllable direction. The navigation policy is parameterized by a neural network and optimized using a genetic algorithm that evolves an ensemble of candidate strategies through Darwinian selection and parameter mutation, without crossover. The method recovers analytical and numerical optimal-control solutions in several benchmark cases, including two-dimensional turbulence, and outperforms Q-learning and one-step actor-critic method. We further show that the learned navigation strategies are robust to variations in the starting position and to unresolved small-scale turbulent fluctuations, provided that the characteristic velocity of these fluctuations remains small compared to the vessel's slip velocity. The proposed approach provides a compact representation of the navigation policy, is readily parallelizable, and eliminates the need for reward shaping, offering an efficient alternative to existing analytical and numerical methods for computing minimum-time trajectories in complex flows.
\end{abstract}

\maketitle

\section{Introduction}

The minimum-time navigation of a vessel with fixed-magnitude slip velocity and controllable direction in a prescribed two-dimensional flow is known as Zermelo’s navigation problem~\cite{zermelo1931navigationsproblem}. Nearly a century after its formulation, it remains a fundamental problem in flow-assisted navigation, with applications ranging from micro-robots for targeted drug and gene delivery~\cite{nelson2010microrobots, esteban2016acoustically, park2017multifunctional, luo2018micro, erkoc2019mobile, manzari2021targeted} to autonomous vehicles and ocean gliders operating in oceanic and atmospheric currents~\cite{trincavelli2008towards, guerrero2013uav, zeng2016comparison, lermusiaux2017future, bellemare2020autonomous, chai2020monitoring, panda2020comprehensive}. The objective is to optimally exploit the interplay between self-propulsion and environmental advection to minimize travel time. In the present formulation, the flow field is assumed to be known a priori and available to the vessel, in contrast to navigation strategies based solely on local sensory information~\cite{colabrese2017flow, gustavsson2017finding, alageshan2020machine, yang2020efficient, gunnarson2021learning, monthiller2022surfing, calascibetta2023optimal, mecanna2025critical}.

The classical optimal-control formulation transforms Zermelo's problem into a two-point boundary-value problem, yielding locally optimal trajectories~\cite{bryson2018applied, biferale2019zermelo, parfenyev2026optimal}. Analytical solutions exist for several simple flow configurations~\cite{techy2011optimal, liebchen2019optimal}, whereas more general flows require numerical integration combined with shooting methods~\cite{kirk2004optimal, bonnans2013shooting, biferale2019zermelo}. Extensions have considered three-dimensional flows~\cite{guerrero2013uav}, navigation on curved manifolds~\cite{piro2021optimal, piro2022efficiency}, and alternative objectives such as fuel consumption and energy dissipation~\cite{liebchen2019optimal, piro2024optimal}. A fundamental difficulty is that the resulting optimal trajectories can exhibit strong sensitivity to initial conditions and controls. The steering degree of freedom makes the system susceptible to chaotic behavior, causing optimal trajectories to become unstable even in regular currents~\cite{biferale2019zermelo, piro2021optimal, piro2022efficiency, parfenyev2026optimal}. Consequently, shooting methods offer no general guarantee of finding the global optimum and yield solutions that are highly sensitive to small perturbations, limiting their robustness in complex flows.

Reinforcement learning (RL) offers an alternative framework in which robust navigation strategies can be learned through repeated interactions with the flow~\cite{sutton1998reinforcement}. In particular, Q-learning has been applied to path planning in steady environments~\cite{yoo2016path, schneider2019optimal}, while a one-step actor-critic method has been used for navigation in more complex two-dimensional turbulent flows~\cite{biferale2019zermelo, buzzicotti2020optimal}. A recent comparison of such tabular RL approaches with optimal-control methods across flows of increasing complexity showed that, although these methods can reliably reach the target, the resulting trajectories are generally suboptimal, with the discrepancy becoming particularly pronounced in time-dependent turbulent settings~\cite{parfenyev2026optimal}. This motivates the use of more expressive deep RL methods for Zermelo's navigation problem. An initial step in this direction was taken in Ref.~\onlinecite{nasiri2022reinforcement}, where an Advantage Actor-Critic (A2C) algorithm reproduced analytical optimal-control solutions for several steady flow configurations. These results demonstrated that deep RL can recover optimal navigation strategies and suggested a route toward problems for which conventional optimal-control methods become challenging.

Here, we develop a complementary approach that combines neural-network representations of navigation policies with population-based optimization using a genetic algorithm~\cite{mitchell2009complexity, such2017deep}. The parameters of the neural network (NN) define the genome of each individual in a population, and individuals are evaluated according to a fitness function based on the travel time. The best-performing policies are selected to form the next generation through mutations of their NN parameters, and the procedure is iterated until the fitness ceases to improve. Unlike gradient-based optimization, this approach does not require differentiability of the objective with respect to the policy parameters. Moreover, fitness evaluations for different individuals are independent and can therefore be performed in parallel, allowing efficient utilization of computational resources.

We validate the approach across a hierarchy of flow configurations with increasing complexity considered in Ref.~\onlinecite{parfenyev2026optimal}. For steady vortex and time-dependent sink-vortex flows, for which analytical optimal-control solutions are available~\cite{techy2011optimal, parfenyev2026optimal}, the learned policies reproduce the corresponding optimal trajectories. 
For both a snapshot of two-dimensional turbulence in the inverse-cascade regime and a fully time-dependent turbulent flow, the resulting trajectories closely follow the numerically computed optimal-control paths, achieving shorter travel times than Q-learning and a one-step actor-critic method, while also exhibiting robustness to variations in the vessel's starting position. For the time-dependent turbulent flow, we further examine the robustness of the learned policies to unresolved small-scale velocity fluctuations that may arise from imperfect flow forecasts. We train the NNs using coarse-grained velocity fields with different filtering thresholds, thereby controlling the quality of the available flow forecast, and evaluate the resulting policies in the full-resolution flow. The learned strategies successfully reach the target provided that the characteristic velocity of the unresolved fluctuations remains sufficiently small compared with the vessel's slip velocity. The obtained results demonstrate that genetic-algorithm optimization of neural-network policies provides a practical framework for learning efficient and robust navigation strategies in complex flows.

\section{Problem Setup}

The motion of a vessel is described by the kinematic equation
\begin{equation}\label{eq:motion}
\dfrac{d \bm R_t}{dt} = \bm v(\bm R_t, t) + V_0
\begin{pmatrix}
\cos \theta_t \\
\sin \theta_t
\end{pmatrix},
\end{equation}
where $\bm R_t = (X_t, Y_t)$ denotes the position of the vessel at time $t$ in Cartesian coordinates, $\bm v(\bm r,t)$ represents the underlying 2D advecting flow, and $V_0$ is the magnitude of the vessel's slip velocity relative to the flow. The steering angle $\theta_t$ acts as the control variable and can be adjusted dynamically throughout the journey. The objective of Zermelo's navigation problem is to determine a control policy $\theta_t=\theta(\bm R_t,t)$ that minimizes the total travel time $T$ required to reach a prescribed destination $\bm r_B$ from a starting position $\bm r_A$.

Using optimal control theory, one can prove that if such trajectory exists, then the steering angle satisfies~\cite{techy2011optimal, parfenyev2026optimal} 
\begin{equation}\label{eq:angle} 
%\begin{split} 
\dot{\theta}_t = \partial_x v_y \sin^2 \theta_t - \partial_y v_x \cos^2 \theta_t %\\
+ (\partial_x v_x-\partial_y v_y) \sin \theta_t \cos \theta_t,
%\end{split} 
\end{equation}
where the velocity gradients are evaluated along the vessel's trajectory $\bm R_t$. The resulting system (\ref{eq:motion})-(\ref{eq:angle}) must be supplemented with boundary conditions $\bm R_0 = \bm r_A$ and $\bm R_T = \bm r_B$. Note that the initial heading $\theta_0$ and travel time $T$ are not specified and therefore remain free parameters that must be determined. 

A standard approach to solve this two-point boundary value problem is the shooting method~\cite{kirk2004optimal, bonnans2013shooting}: an initial guess for $\theta_0$ is refined until the resulting trajectory, obtained by forward integration of the system, satisfies the terminal condition $\bm R_T = \bm r_B$ within a desired tolerance. Importantly, the considered system may admit multiple solutions: Zermelo's equation (\ref{eq:angle}) provides a necessary condition for a local extremum but does not guarantee a global optimum~\cite{biferale2019zermelo}. Moreover, the resulting trajectories can be unstable and characterized by positive finite-time Lyapunov exponents. In these cases, infinitesimal uncertainties in starting positions or control inputs are exponentially amplified, leading to drastic deviations from the optimal route~\cite{biferale2019zermelo, piro2021optimal, piro2022efficiency, parfenyev2026optimal}. Consequently, shooting methods provide no general guarantee of finding the global optimum and the practical application of optimal control is severely limited. For steady flows, one can apply heuristic algorithms on top of unstable optimal trajectories to compensate for deviations, thereby enabling more robust navigation~\cite{piro2022optimal}.

\section{Genetic Algorithm}

\begin{figure}[t]
    \centering
    \includegraphics[width=0.85\linewidth]{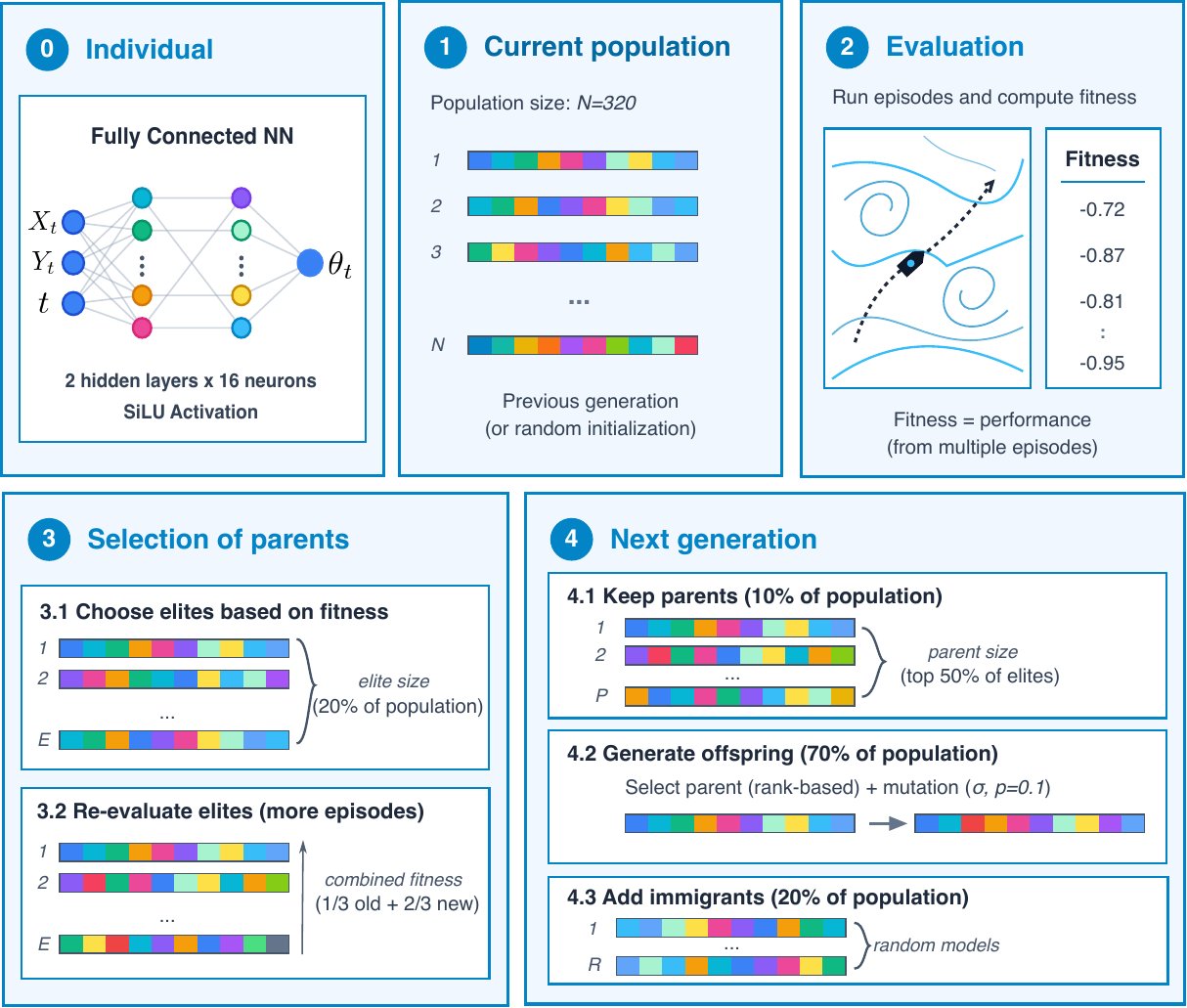}
    \caption{Schematic of the genetic algorithm. Box $0$ depicts the neural network encoding an individual’s navigation strategy, with its parameters defining the genome. Boxes $1$–$4$ illustrate the successive stages of generating a new generation from the preceding one. The process is repeated until fitness no longer improves.}
    \label{fig:1}
\end{figure}

We propose a complementary approach to Zermelo’s navigation problem that learns robust navigation strategies through repeated interactions with the flow. An overview of the algorithm is provided in Fig.~\ref{fig:1}, followed by a detailed description below.

The navigation strategy is represented by a fully connected NN with $N_l=2$ hidden layers, each containing $N_n=16$ neurons with SiLU activation~\cite{elfwing2018sigmoid}. The NN takes the system state as input: the vessel position $(X_t,Y_t)$ for steady flows and the vessel position and time $(X_t,Y_t,t)$ for time-dependent flows. Its single linear output represents the steering direction $\theta_t$. Unlike previous tabular RL approaches~\cite{biferale2019zermelo, buzzicotti2020optimal, parfenyev2026optimal}, which define the system state solely by the vessel position and thus effectively formulate navigation in time-dependent flows as a partially observable Markov decision process to avoid an excessively large state space, our approach explicitly includes time in the system state. Moreover, the proposed strategy representation has a highly compact parameterization, requiring substantially fewer trainable parameters than previous approaches~\cite{schneider2019optimal, biferale2019zermelo, buzzicotti2020optimal, parfenyev2026optimal, nasiri2022reinforcement}, while operating directly in continuous state and action spaces without discretization.

The strategy is evaluated over episodes, each corresponding to an attempt to reach the target. The initial position $\bm R_0$ is sampled uniformly within a disk of radius $d_A$ centered at $\bm r_A$, thereby promoting robustness to variations in the initial condition. The heading $\theta_t$ is updated according to the NN prediction at discrete intervals $\Delta t$ and held fixed between updates, while Eq.~(\ref{eq:motion}) is numerically integrated. The interval $\Delta t$ is sufficiently small, with the resulting minimum-time trajectories being two orders of magnitude longer. An episode terminates when the vessel reaches a circle of radius $d_B$ centered at $\bm r_B$, the maximum duration $T_{\max}$ is exceeded, or the vessel collides with the domain boundary. The value of $T_{\max}$ is chosen to be several times the typical travel time, allowing sufficient time to explore the flow while terminating episodes in which the vessel becomes trapped. The strategy performance is quantified by the fitness, defined as the negative episode duration, with boundary collisions incurring an additional strong penalty of $-2T_{\max}$. Compared to tabular RL methods~\cite{biferale2019zermelo, buzzicotti2020optimal, parfenyev2026optimal}, no distance-based potential is required in the reward. The introduced parameters define the navigation problem independently of the optimization procedure. For the turbulent flows considered below, $\bm r_A$, $\bm r_B$, $d_A$, $d_B$, and $\Delta t$ are identical to those used in Ref.~\onlinecite{parfenyev2026optimal}, facilitating direct comparison.

To optimize the navigation strategy, we employ a genetic algorithm~\cite{mitchell2009complexity, such2017deep}, in which a population of $N=320$ randomly initialized strategies evolves over successive generations through selection and parameter mutation. In each generation, the population is evaluated over $n_{ep}$ episodes and ranked according to the mean fitness. The top 20\% of individuals are designated as elites and reevaluated over an additional $2n_{ep}$ episodes to reduce statistical fluctuations in the fitness estimate. The top 50\% of elites are selected as parents and copied unchanged to the next generation, corresponding to 10\% of the population.

The remaining 70\% of the population is generated by randomly selecting a parent according to its rank and mutating its NN parameters. Each parameter is mutated with probability $p_{\sigma}=0.1$ by adding Gaussian noise with zero mean and variance $\sigma^2$. We do not employ crossover, since the neurons within each hidden layer are permutation symmetric, such that corresponding neurons in different networks need not represent the same behavioral features. The final 20\% of the population consists of randomly initialized individuals, referred to as immigrants. These individuals enhance exploration of the strategy space and help prevent premature convergence to local optima.

After a new generation is formed, the evaluation, selection, and reproduction steps are repeated. The evolution is continued for $N_{gen}=200$ generations, which is sufficient for the best fitness to reach a plateau. To balance exploration and exploitation, the number $n_{ep}$ of evaluation episodes and mutation amplitude $\sigma$ are varied throughout the evolution. Early generations use fewer episodes and larger mutations, providing inexpensive and coarse fitness estimates while promoting exploration. As the population improves, the fitness is evaluated more accurately and the mutation amplitude is reduced to enable finer parameter optimization. Specifically, $n_{ep} = \min (10, 2 \cdot 1.02^{\text{idx}})$ and $\sigma = 0.5 \cdot 0.99^{\text{idx}}$, where $\text{idx}$ denotes the generation index. 

The final navigation strategy is taken as the highest-fitness individual in the last generation. Importantly, all hyperparameters of the genetic algorithm are kept identical for all flow examples considered below. This consistency indicates that the proposed approach is relatively insensitive to their variation across the different flow environments. Finally, we emphasize that the outcome of the genetic algorithm is a navigation policy $\theta(\bm R_t, t)$, which maps possible system states onto actions -- selection of steering angle. In contrast, the optimal-control approach yields the dependence $\theta(\bm R_t^*,t)$ associated with a single optimal trajectory $\bm R^*_t$.

\section{Results}

To evaluate the resulting navigation strategy, we perform $n_{eval} = 5000$ evaluation episodes with perturbed initial positions $\bm r_A$, as described above. We quantify performance using the success probability $p$, defined as the fraction of trajectories that reach the target, and the mean travel time $\langle T_G \rangle$ computed over successful trajectories. For navigation in turbulent flows, these performance metrics may exhibit variability arising from the stochasticity of the training process, particularly when the flow forecast is coarse-grained. We therefore also report their standard deviations, estimated across $10$ independent training runs.

\subsection{Analytical examples}

\begin{figure}[b]
    \centering
    \includegraphics[width=0.6\linewidth]{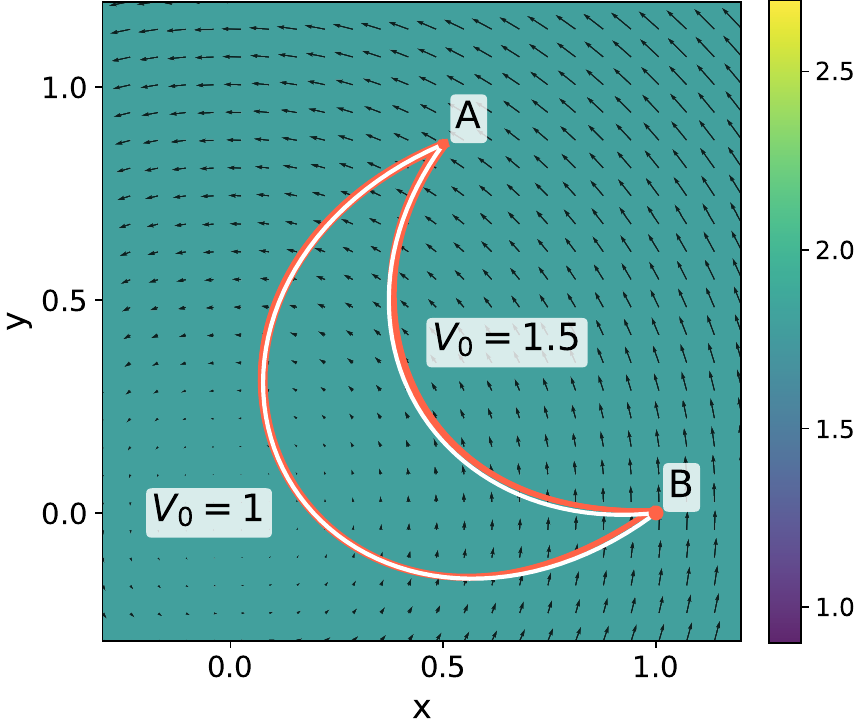}
    \caption{Optimal-control solutions (white) and trajectories obtained via the genetic algorithm (red) for the steady vortex flow (\ref{eq:sv}). The parameters are $\omega=0.9$, $\bm r_A = (1/2,\sqrt{3}/2)$, $\bm r_B = (1,0)$, $V_0=1$ and $1.5$. For the genetic algorithm, we use $d_A=0.01$, $d_B=0.015$, $\Delta t=0.012$, $T_{\max}=5$.}
    \label{fig:2}
\end{figure}

We first consider navigation in the steady vortex flow
\begin{equation}\label{eq:sv}
    \bm v(x,y) = \left(-\omega y, \; \omega x \right),
\end{equation}
corresponding to solid-body rotation, see Fig.~\ref{fig:2}. The optimal-control equations (\ref{eq:motion})--(\ref{eq:angle}) can be integrated analytically, yielding the optimal trajectories connecting $\bm r_A$ and $\bm r_B$~\cite{techy2011optimal, parfenyev2026optimal}. Perturbations in either the initial position or the control input grow linearly rather than exponentially in time, indicating that these optimal trajectories are only weakly unstable~\cite{parfenyev2026optimal}. The trajectories obtained with the policies learned by the genetic algorithm closely reproduce the analytical optimal-control solutions. All evaluated trajectories successfully reach the target, with mean travel times of $\langle T_G\rangle\approx1.964$ for $V_0=1$ and $\langle T_G\rangle\approx1.144$ for $V_0=1.5$. In both cases, the measured travel times are slightly shorter than the analytical results, $T \approx 1.975$ for $V_0=1$ and $T \approx 1.151$ for $V_0=1.5$, due to the finite target size.

\begin{figure}[t]
    \centering
    \includegraphics[width=0.6\linewidth]{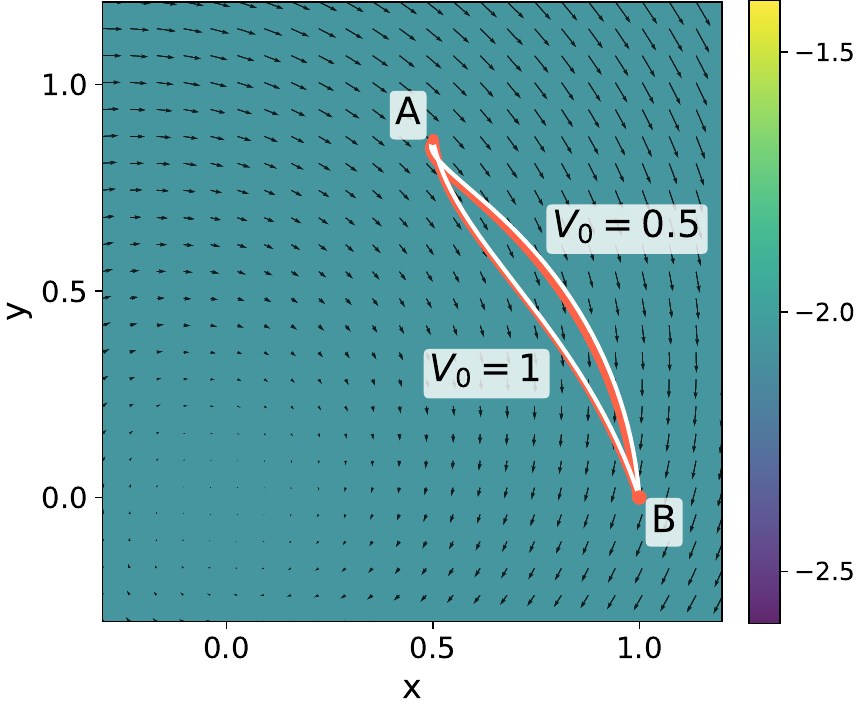}
    \caption{Optimal-control solutions (white) and trajectories obtained via the genetic algorithm (red) for the time-varying sink-vortex flow (\ref{eq:sink+rotation}). The parameters are $\bm r_A = (1/2,\sqrt{3}/2)$, $\bm r_B = (1,0)$, $V_0=1$ and $0.5$. For the genetic algorithm, we use $d_A=0.01$, $d_B=0.015$, $\Delta t=0.012$, $T_{\max}=5$.}
    \label{fig:3}
\end{figure}

Next, we consider navigation in a time-dependent flow consisting of a steady sink and a vortex with vorticity that varies linearly with time
\begin{equation}\label{eq:sink+rotation}
    \bm v (x,y, t) =
    \begin{pmatrix}
    -0.3 x - (0.5-t)y \\
    -0.3 y + (0.5-t)x
    \end{pmatrix}.
\end{equation}
In this case, the optimal-control solutions can also be obtained analytically~\cite{techy2011optimal, parfenyev2026optimal} and they are neutrally stable, with perturbations in the initial conditions remaining bounded in time~\cite{parfenyev2026optimal}. As shown in Fig.~\ref{fig:3}, the policies learned by the genetic algorithm again produce trajectories that closely reproduce the analytical solutions for the specified parameters. All evaluated trajectories successfully reach the target, with mean travel times of $\langle T_G\rangle\approx1.023$ for $V_0=1$ and $\langle T_G\rangle\approx 1.530$ for $V_0=0.5$, again slightly shorter than the analytical results, $T \approx 1.030$ for $V_0=1$ and $T \approx 1.540$ for $V_0=0.5$. Taken together, these examples demonstrate that the navigation strategies learned by the genetic algorithm accurately recover the optimal trajectories in both steady and time-dependent flows. This agreement validates the approach and motivates its application to more complex flows where analytical solutions are unavailable.

\subsection{2D Turbulence}

\begin{figure}[t]
    \centering
    \includegraphics[width=0.65\linewidth]{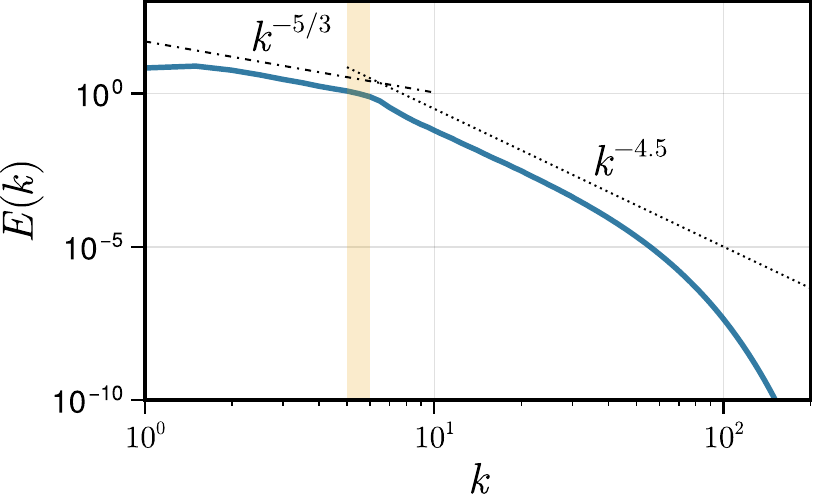}
    \caption{Energy spectrum for DNS of 2D turbulence.}
    \label{fig:4}
\end{figure}

In all other examples considered in this work, the background velocity $\bm v(\bm r,t)$ is obtained from a statistically stationary solution of the 2D Navier–Stokes equations,
\begin{equation}\label{eq}
\partial_t \bm v + (\bm v \cdot \nabla) \bm v = - \nabla p + \nu \nabla^2 \bm v - \alpha \bm v + \bm f, \quad \nabla \cdot \bm v = 0,
\end{equation}
computed in a periodic square box of size $L=2\pi$ using a standard pseudospectral solver~\cite{GeophysicalFlowsJOSS}. The flow is maintained by stochastic forcing $\bm f$ centered at the wavenumber $k_f = 5.5$. The resulting turbulent state is characterized by the dimensionless parameters $Re = UL/\nu \approx 8 \cdot 10^3$ and $Rh = U/(\alpha L) \approx 5$, where $U \approx 6.3$ is the root-mean-square velocity, $\nu = 5 \cdot 10^{-3}$ is the kinematic viscosity and $\alpha = 0.2$ is the bottom-drag coefficient. The mean energy spectrum, shown in Fig.~\ref{fig:4}, is consistent with the established phenomenology of forced 2D turbulence. In the inverse energy cascade range $k<k_f$, the spectrum exhibits the theoretical $E(k) \propto k^{-5/3}$ scaling~\cite{boffetta2012two}, while in the direct enstrophy cascade range $k>k_f$, the presence of bottom drag results in a steeper spectrum than the classical $E(k) \propto k^{-3}$ prediction~\cite{valadao2025spectrum}. Additional details on the DNS setup can be found in Ref.~\onlinecite{parfenyev2026optimal}. 

Unless otherwise stated, the vessel's slip speed is set to $V_0 \approx 4.5$, corresponding to 30\% of the maximum flow velocity. First, we consider navigation in an instantaneous snapshot of 2D turbulence and examine two different sets of starting and ending points, denoted by P1 and P2. The considered examples exactly coincide with those studied in Ref.~\onlinecite{parfenyev2026optimal}, allowing us to compare the performance of the genetic algorithm with predictions from optimal control theory and RL methods. Note that in turbulent flows, the optimal-control boundary-value problem has multiple solutions, which are unstable and therefore difficult to find using shooting methods~\cite{biferale2019zermelo, parfenyev2026optimal}. Consequently, the global optimality of the reported minimal-time optimal-control solutions cannot be rigorously established, although the close agreement with independent navigation methods provides supporting evidence.

\begin{figure}[b]
    \centering
    \includegraphics[width=\linewidth]{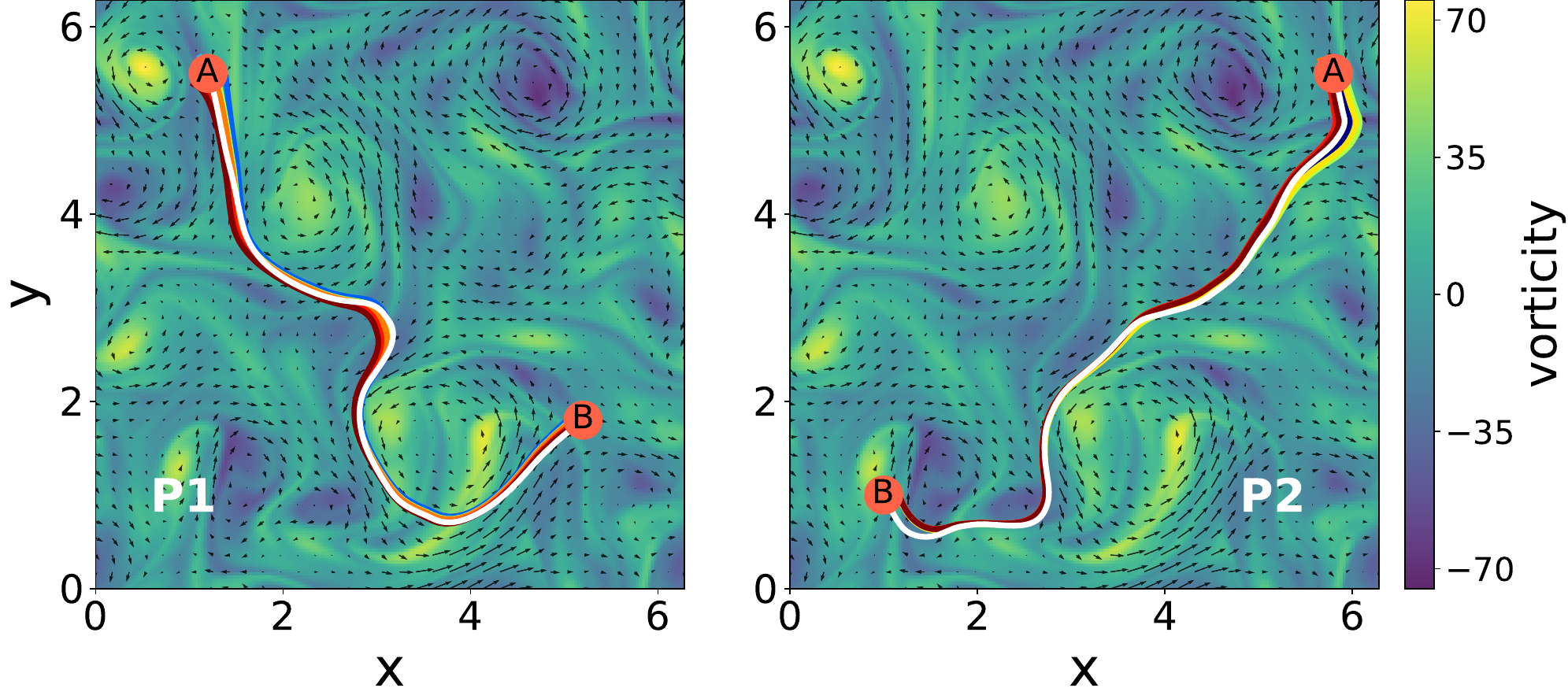}
    \caption{Optimal-control solutions (white) and several trajectories with random initial positions obtained via the genetic algorithm (other colors) for the turbulence snapshot. The vessel's slip speed is $V_0 \approx 4.5$. For the genetic algorithm, we use $d_A=d_B=0.2$, $\Delta t \approx 0.007$, and $T_{\max}=4$.}
    \label{fig:5}
\end{figure}

Figure~\ref{fig:5} compares the trajectories obtained using the policies learned by the genetic algorithm with the corresponding optimal-control solutions~\cite{parfenyev2026optimal}. The trajectories are in close agreement, with minor discrepancies attributable to the finite sizes $d_A=d_B=0.2$ of the initial and target regions. The optimal-control solutions shown here and throughout the paper correspond to point-to-point navigation from $\bm r_A$ to $\bm r_B$. 

To assess the stochastic variability of the training process, we perform $10$ independent training runs. For both P1 and P2 problems, all runs achieve a success probability of $p=1$, indicating robust navigation under variations in the vessel's initial position. Table~\ref{tab:1} summarizes the mean travel times and standard deviations, together with the optimal-control times and the corresponding results for Q-learning and the one-step actor-critic method from Ref.~\onlinecite{parfenyev2026optimal}. As in the analytical examples above, the genetic algorithm achieves travel times slightly shorter than the point-to-point optimal-control times, owing to the finite sizes of the initial and target regions. For both P1 and P2 problems, it also substantially outperforms both RL methods in travel time and exhibits smaller run-to-run variations, indicating greater training stability.

\begin{table}[t]
\caption{Travel times for optimal-control trajectories ($T$), genetic algorithm ($T_G$), Q-learning ($T_Q$), and one-step actor-critic method ($T_{AC}$). Here, "S" denotes a turbulence snapshot and "T" a time-dependent turbulent flow. The optimal-control times correspond to point-to-point navigation, whereas the other methods use initial and target regions with radii $d_A=d_B=0.2$.}
\begin{ruledtabular}
\begin{tabular}{c|c|c|c|c}
 & $T$ & $\langle T_G \rangle$ & $\langle T_Q \rangle$ & $\langle T_{AC} \rangle$\\
\hline
2D turb (S), P1   & $0.820$ & $0.816 \pm 0.002$ & $0.94 \pm 0.03$ & $0.87 \pm 0.005$\\
2D turb (S), P2   & $0.820$ & $0.801 \pm 0.002$ & $0.99 \pm 0.05$ & $1.03 \pm 0.01$\\
2D turb (T), P3   & $0.670$ & $0.688 \pm 0.001$ & $0.90 \pm 0.06$ & $1.17 \pm 0.04$\\
2D turb (T), P4   & $0.700$ & $0.674 \pm 0.001$ & $1.04 \pm 0.05$ & $1.00 \pm 0.03$\\
\end{tabular}
\end{ruledtabular}
\label{tab:1}
\end{table}

\begin{figure}[b]
    \centering
    \includegraphics[width=0.9\linewidth]{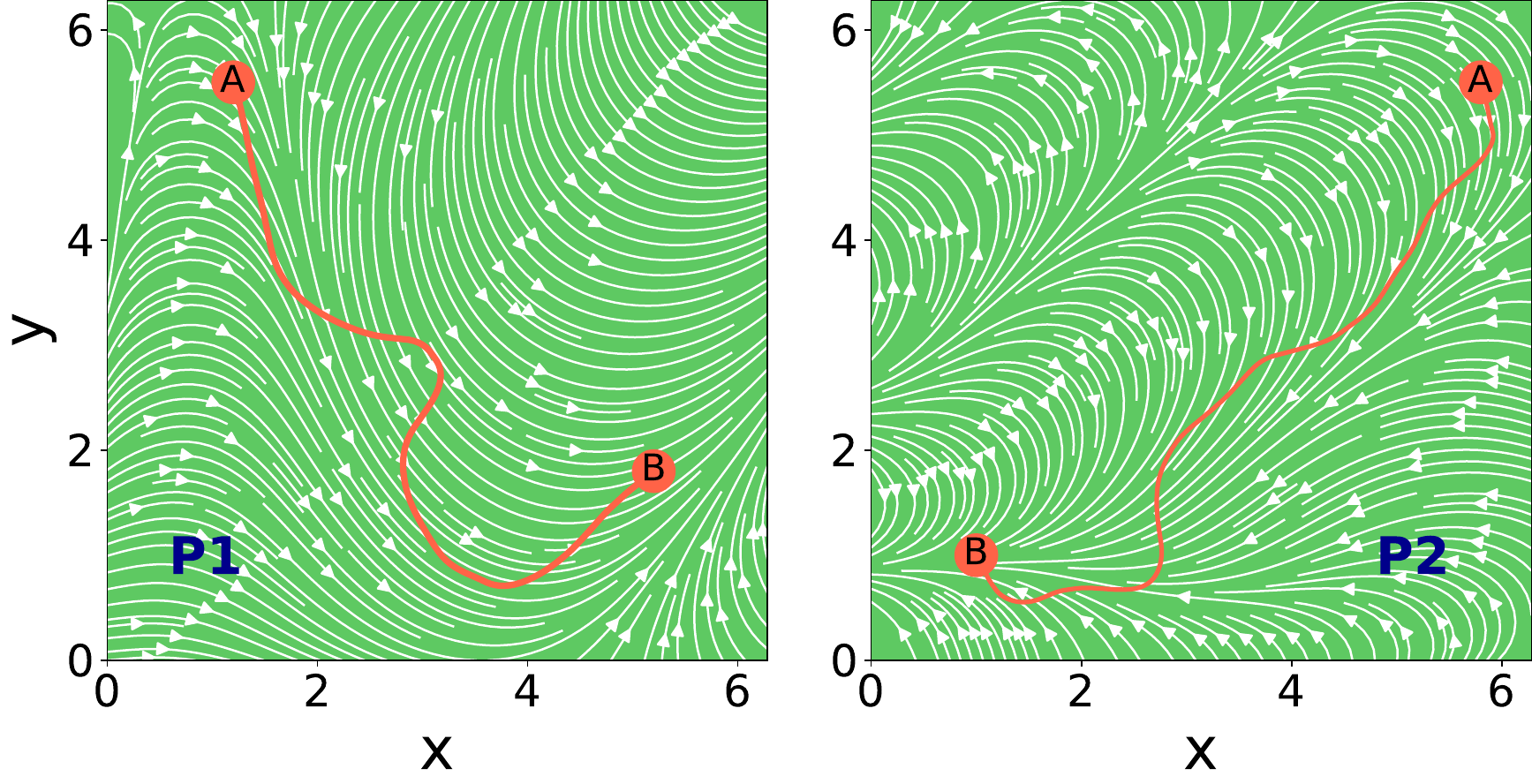}
    \caption{Streamplots of the steering direction $\theta(x,y)$ learned by the genetic algorithm for the turbulence snapshot problems. The optimal-control trajectories are shown for reference.}
    \label{fig:6}
\end{figure}

Figure~\ref{fig:6} illustrates the navigation policies learned by the genetic algorithm. The white streamlines indicate the local direction $\theta(x,y)$ of the self-propulsion velocity as a function of vessel position. Because training trajectories are concentrated near dynamically relevant regions, the learned policy is expected to be most accurate in the vicinity of the optimal trajectories, whereas regions visited less frequently during training may exhibit suboptimal behavior. This spatially nonuniform allocation of computational effort enables the algorithm to focus resources on the most relevant regions of phase space, in contrast to conventional methods of dynamic programming.

\begin{figure}[t]
    \centering
    \includegraphics[width=\linewidth]{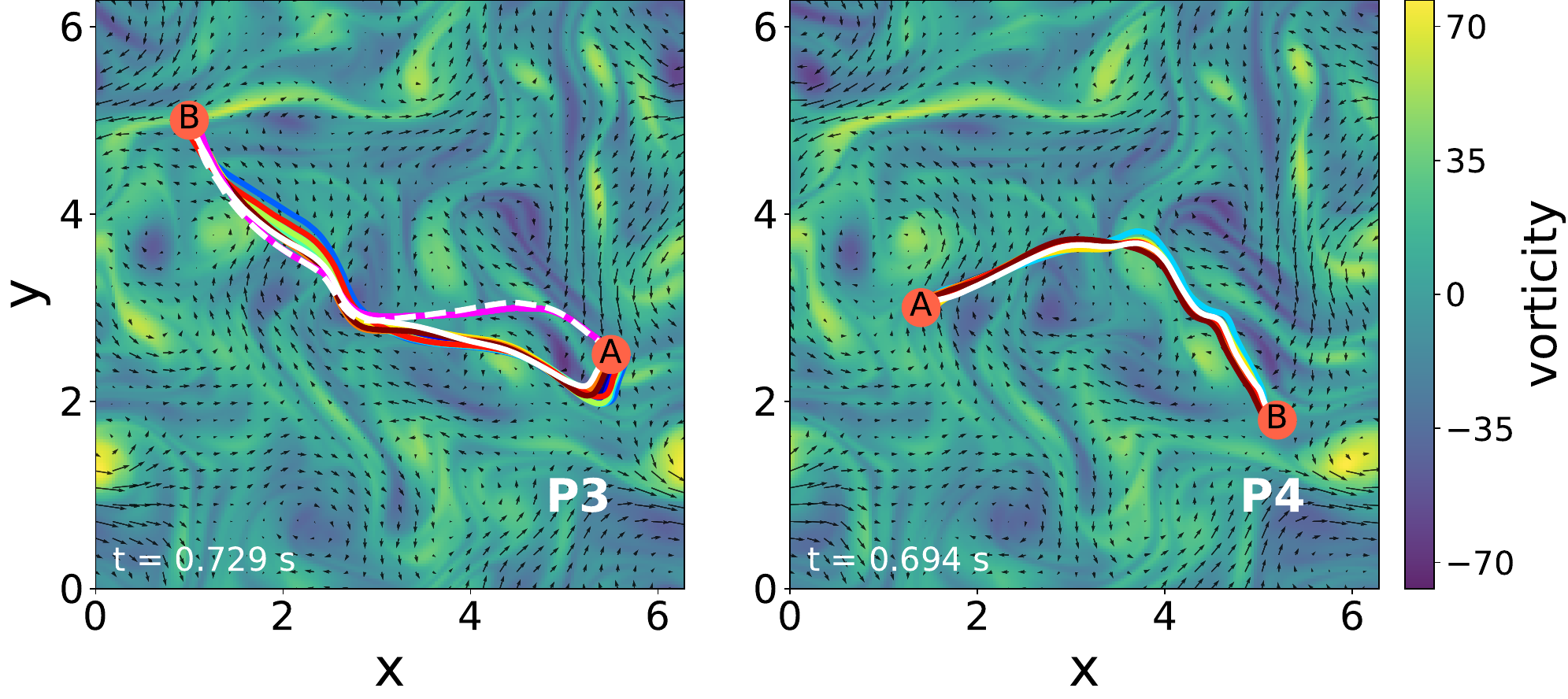}
    \caption{Trajectories generated by the learned navigation strategies for the P3 and P4 problems in a time-dependent turbulent flow. White lines show the corresponding optimal-control solutions. For the P3 problem, the solid and dashed white lines denote the solution for $d_A=0.2$ and another optimal-control solution with shorter travel time, approached as $d_A\to0$, respectively. All other parameters are as in Fig.~\ref{fig:5}.}
    \label{fig:7}
\end{figure}

Next, we consider navigation in a time-dependent turbulent flow for two sets of initial and target points, denoted P3 and P4. Figure~\ref{fig:7} compares the trajectories generated by the navigation strategies learned with the genetic algorithm with the corresponding optimal-control solutions~\cite{parfenyev2026optimal}; a video demonstration is provided in the Supplemental Material~\cite{SM}. For the P3 problem, the learned navigation strategy generates trajectories that converge to a locally optimal solution (white solid line), which results from the finite size $d_A=0.2$ of the starting region. Retraining the algorithm with decreasing $d_A$ shows that, in the limit $d_A\to0$, the learned strategy instead generates trajectories converging to another optimal-control solution with a shorter travel time (white dashed line). Thus, the observed suboptimality reflects a trade-off between travel-time minimization and robustness to variations in the starting position. For the P4 problem, as in the turbulence-snapshot case, the learned navigation strategy generates trajectories that closely follow the identified minimum-time optimal-control solution.

The success probability, estimated over $10$ independent training runs, is $p=0.98\pm0.03$ for the P3 problem and $p=1$ for the P4 problem. The corresponding mean travel times and standard deviations are reported in Table~\ref{tab:1}. For the P3 problem, the genetic algorithm yields a slightly longer mean travel time than the optimal-control solution shown by the dashed line, reflecting the suboptimality discussed above; for the solid-line solution, $T\approx0.690$. For the P4 problem, the mean travel time is again shorter than the point-to-point optimal-control time because of the finite sizes of the initial and target regions. Notably, the genetic algorithm substantially outperforms the tabular RL methods of Ref.~\onlinecite{parfenyev2026optimal}, yielding markedly shorter mean travel times and smaller run-to-run variability. The travel times for both P3 and P4 problems are of the same order as the turnover time of system-size eddies $L/U\sim1$, indicating that the vessel experiences substantial flow evolution during navigation. These results demonstrate that the performance gains observed for turbulence snapshots persist, and are even more pronounced, in time-dependent turbulent flows.

\subsection{Coarse-grained forecast}

Finally, we consider navigation when the vessel has access only to coarse-grained information about the external turbulent flow, as relevant to path planning in oceanic and atmospheric currents. We revisit the P3 problem in the time-dependent turbulent flow introduced above. The DNS velocity field is filtered by removing Fourier modes with wavenumbers $k>k_{\max}$, and the genetic algorithm is applied using the resulting coarse-grained flow. We consider $k_{\max}=\{256,10,5,3,2\}$, where $k_{\max}=256$ corresponds to the fully resolved DNS field. The corresponding root-mean-square velocity of the unresolved fluctuations is $V_{fluct}\approx \{0,0.6,2.0,3.3,4.3\}$, respectively. For each filtering level, we also vary the vessel slip velocity as $V_0\approx \{5.7,3.8,2.5,1.9\}$. After training, each navigation strategy is evaluated in the fully resolved velocity field. We assess whether the vessel can reach the target despite unresolved velocity fluctuations and quantify the resulting travel time. To reduce stochastic variability, the success probability $p$ and mean travel time of successful trajectories $\langle T_G\rangle$ are averaged over $10$ independent training runs.

\begin{figure}[t]
    \centering
    \includegraphics[width=\linewidth]{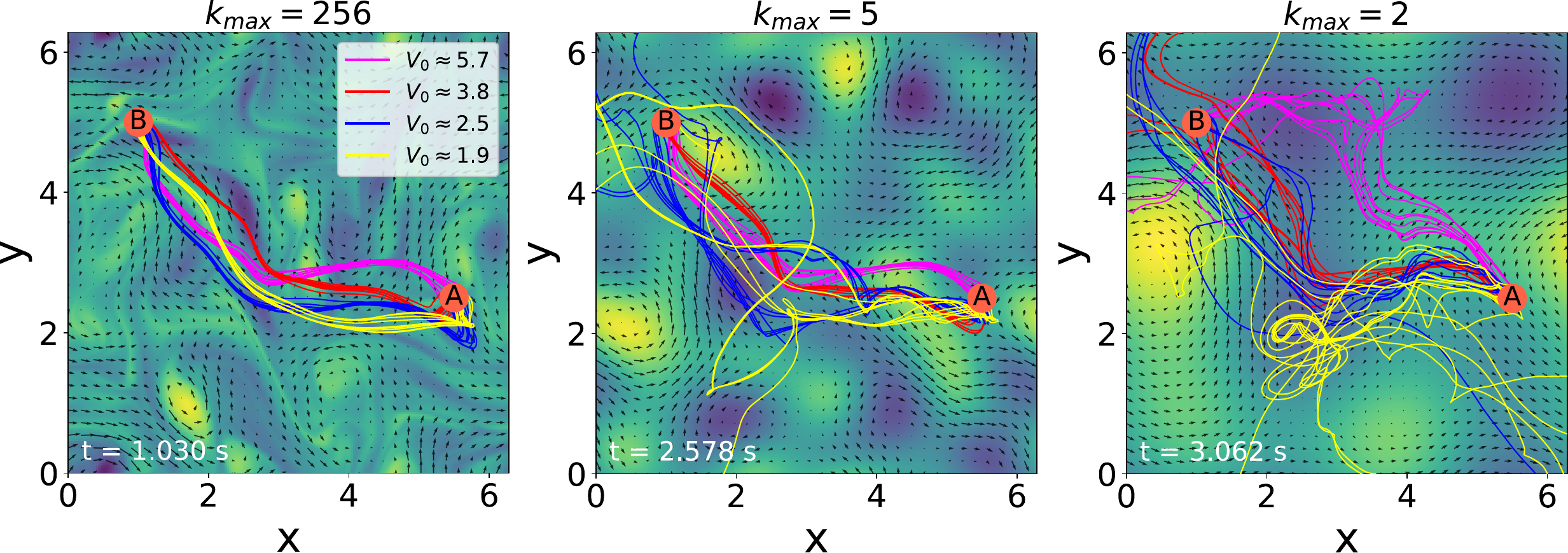}
    \caption{Representative trajectories generated by the navigation strategies learned with the genetic algorithm for different levels $k_{max}$ of flow-field filtering and vessel self-propulsion speed $V_0$.}
    \label{fig:8}
\end{figure}

Figure~\ref{fig:8} shows representative trajectories generated by the learned policies. Under a perfect flow forecast, trajectories originating from nearby initial positions remain close to one another forming bunches, and nearly all reach the target. The optimal routes, however, depend strongly on the vessel's self-propulsion speed $V_0$. As the forecast quality deteriorates, unresolved velocity fluctuations increasingly distort the trajectories, causing initially nearby paths to diverge substantially. This effect becomes more pronounced as the vessel speed $V_0$ decreases. We also observe an increase in run-to-run variability with deteriorating forecast quality.

\begin{figure}[b]
    \centering
    \includegraphics[width=\linewidth]{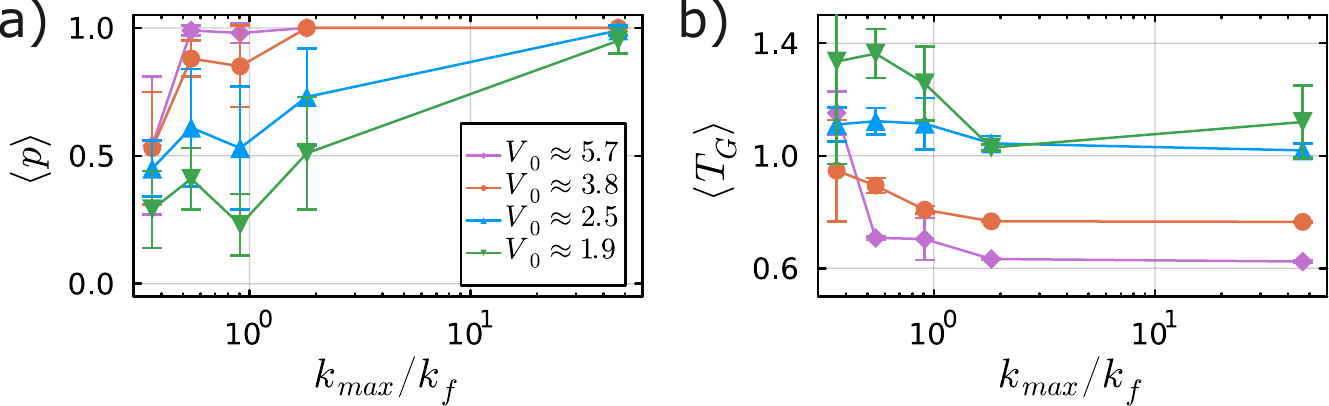}
    \caption{Navigation performance as a function of normalized filtering scale $k_{max}/k_f$ and vessel self-propulsion speed $V_0$. (a) Probability of reaching the target $\langle p \rangle$ and (b) mean travel time of successful trajectories $\langle T_G\rangle$ evaluated in the fully resolved turbulent flow. Error bars show the standard deviation.}
    \label{fig:9}
\end{figure}

This behavior is quantified in Fig.~\ref{fig:9}. Panel (a) shows that the target-reaching probability decreases with increasing filtering of the velocity field, with the curves for smaller $V_0$ being more sensitive to forecast quality. When the characteristic velocity of the unresolved fluctuations is small compared with $V_0$, their effect is weak; once the two become comparable, the trajectories are substantially distorted. This distortion is accompanied by an increase in the mean travel time to the target, shown in panel (b). A qualitatively similar behavior was reported for tabular RL methods in Ref.~\onlinecite{parfenyev2026optimal}. Here, we extend this analysis to the genetic algorithm and demonstrate the dependence on the vessel speed $V_0$. Overall, the genetic algorithm produces navigation strategies that remain robust to noise from unresolved turbulent fluctuations as long as their characteristic amplitude is small compared with the vessel's self-propulsion speed $V_0$.

\section{Conclusion}

We have demonstrated that optimizing a neural-network navigation policy via a genetic algorithm provides an effective and robust solution to Zermelo's navigation problem in two-dimensional flows. The proposed method successfully recovers optimal-control trajectories in both steady and time-dependent turbulent flows, and achieves shorter travel times and lower run-to-run variability than the tabular RL approaches such as Q-learning and one-step actor-critic method. Additionally, the learned navigation strategies exhibit strong robustness to initial position variations and imperfect flow forecasts, provided that unresolved turbulent velocity fluctuations remain smaller than the vessel's slip velocity. These results highlight the potential of genetic optimization as a compact, parallelizable, and broadly applicable method for learning minimum-time navigation strategies in complex flows.

\acknowledgments
The development of the genetic algorithm was supported by the Ministry of Science and Higher Education of the Russian Federation (State Assignment FFWR-2024-0014 of the Landau Institute for Theoretical Physics), while its application to examples was supported by the Basic research program of HSE University (project HSE-BR-2025-57). The author gratefully acknowledges the Landau Institute for providing computing resources.

\section*{Data AVAILABILITY}
The code that support the findings of this study are openly available at \url{https://github.com/parfenyev/navig-ga/}.

\bibliography{biblio}

\end{document}